\documentclass[12pt,preprint]{aastex}

\newcommand{\eps}{{\varepsilon}}

\def\ie         {{\it i.e.,~}}
\def\vs         {{\it vs.~}}
\def\etc         {{\it etc.~}}
\def\etal         {{\it et al.~}}
\def\citeN	{\cite}
\newcommand{\V}[1]{{\bf{#1}}}
\keywords{outer solar system, dynamics, chaos}

\begin{document}

\title{Surfing on the Edge:\\ Chaos \vs Near-Integrability in the System of Jovian Planets}
\author{Wayne B. Hayes}
\affil{Computer Science Department University of California, Irvine\\Irvine, California 92697-3435}
\email{wayne@ics.uci.edu}

\begin{abstract}
We demonstrate that the system of Jovian planets
(Sun+Jupiter+Saturn+Uranus+Neptune), integrated for 200 million
years as an isolated 5-body system using many sets of initial conditions
all within the uncertainty bounds of their currently known positions, can
display both chaos and near-integrability.  The conclusion is consistent
across four different integrators, including several comparisons against
integrations utilizing quadruple precision.  We demonstrate that
the Wisdom-Holman symplectic map using simple symplectic correctors
as implemented in {\it Mercury 6.2} \citep{ChambersMercury99}
gives a reliable characterization
of the existence of chaos for a particular initial condition only
with timesteps less than about 10 days, corresponding to about 400 steps
per orbit.  We also integrate the canonical DE405 initial condition out
to 5 Gy, and show that it has a Lyapunov Time of 200--400 My, opening
the remote possibility of accurate prediction of the Jovian planetary
positions for 5 Gy.
\end{abstract}

\section{Introduction}

\begin{flushright}
{\it Both the man of science and the man of art live always at the edge\\
of mystery, surrounded by it. Both, as a measure of their creation, have\\
always had to do with the harmonization of what is new with what is familiar,\\
with the balance between novelty and synthesis, with the struggle to make\\
partial order in total chaos.... This cannot be an easy life.}\\
--- J. Robert Oppenheimer
\end{flushright}

When one speaks of the stability of our Solar System, one must
carefully define the meaning of ``stable''.  We say that the Solar
System is {\it practically stable} if, barring interlopers, the known
planets suffer no close encounters between themselves or the Sun, over
the main-sequence lifetime of the Sun.  In a practically stable Solar
System, the orbital eccentricities, inclinations, and semi-major axes
of all the planets remain within some bounded region, not too far from
their present values.  In this sense, work by many authors over the past
15 years has all but proven that the Solar System is practically stable
\citep{Laskar94,Laskar96,Laskar97,ItoTanikawa02}.   Good reviews exist
\citep{Lissauer99,LecarFranklinHolmanMurray01}, and we will not discuss
it further in this paper.  A second, more formal definition
involves the question of whether the Solar System is {\it chaotic} or not.
In a chaotic system, nearby solutions tend to diverge from each other
exponentially with time, although in a weakly chaotic system such as
the Solar System, the exponential divergence can be preceded by an initial
period of polynomial divergence.  Let $d(t)$ be the distance between two
solutions, with $d(0)$ being their initial separation.
Then $d(t)$ increases approximately as $d(0) e^{\lambda t}$ in a chaotic
system, where $\lambda$ is the {\it Lyapunov exponent}.  The inverse of
the Lyapunov exponent, $1/\lambda$, is called the {\it Lyapunov time},
and measures how long it takes two nearby solutions to diverge by a
factor of $e$.  A system that is not chaotic is called {\it integrable}
or {\it regular}, and has a Lyapunov exponent of zero.  A practical
consequence of being chaotic is that small changes become exponentially
magnified, so that uncertainties in the current positions of the planets
are magnified exponentially with time.  Even though the Solar System is
practically stable, a positive Lyapunov exponent means that uncertainties
in the current positions of the planets are magnified to the point that
we cannot predict the precise positions of the planets in their orbits
after a few (or at most a few tens of) Lyapunov times.

KAM theory tells us that essentially all Hamiltonian systems which
are not integrable are chaotic.  An initial condition (IC) not lying
precisely on a KAM torus will eventually admit chaos, but with a
time scale that depends critically on the IC.  Symplectic integrators
\citep{ChannellScovel90,WisdomHolman91,SanzSernaActaNum92} have many
nice properties when used for long-term integrations of Hamiltonian
systems, such as conservation of phase-space volume, and bounded energy
error.  However, the validity of symplectically-integrated numerical
solutions also depends critically upon the integration time step $h$,
with the longevity of the solution's validity scaling as $e^{a/h}$
for some constant $a$ \citep{BenettinGiorgilli94,Reich99}.  For linear
problems, the dependence is even stronger and manifests itself as a
bifurcation in the Lyapunov exponent, going discontinuously from zero to
a non-zero value (\citeN{Lessnick96}, \citeN{NewmanLee05} --- but see
\citeN{RauchHolman99}).  Since the solar system is not integrable, and
experiences unpredictable small perturbations, it cannot lie permanently
on a KAM torus, and is thus chaotic.  The operative question is the
time scale of the chaos.  To compute the time scale accurately, we
must guarantee that the measured time scale is not an artifact of the
integration method.

What is the Lyapunov time of the Solar System? \citeN{SussmanWisdom88}
first demonstrated that the motion of Pluto is chaotic with a
Lyapunov time of about 20 million years, corroborated over a longer
integration later by \citeN{KinoshitaNakai96}.  \citeN{Laskar89} performed
an averaged integration of the 8 major planets (excluding Pluto)
and found that the Lyapunov time was about 5 million years, with the
divergence dominated by that of the inner planets.  \citeN{Laskar90}
believed that secular resonances are the cause of the
chaos in the inner Solar System (but see \citeN{MurrayHolman01}),
although he did not believe the system
of Jovian planets was affected by the chaos displayed by the inner planets.
\citeN{SussmanWisdom92} performed a full (non-averaged) integration of the
entire Solar System and confirmed Laskar's 5 million year Lyapunov time,
and further found that the system of Jovian planets by itself had a Lyapunov
time of between 7 and 20 million years, although their measurement of
the Lyapunov time displayed a disturbing dependence on the timestep
of the integration.  This dependence was later discovered to be due
to symplectic integration schemes effectively integrating a slightly
different set of ICs; the effect can be
corrected \citep{SahaTremaine92,WisdomHolmanTouma96}, although it decreases with
decreasing timestep.

Since there are no two-body resonances amongst the Jovian
planets, the cause of the chaos between them was not understood until
\citeN{MurrayHolman99} identified the cause as being the overlap
of three-body resonances.
\citeN{MurrayHolman99} also performed Lyapunov time measurements
on a large set of Outer Solar Systems differing only in the
initial semi-major axis of Uranus.  They found that their three-body
resonance theory correctly predicted which regions of ICs
were chaotic, and which were not, at least over the
200-million-year integration timespan they used.  For the ``actual''
Solar System, they found that the Lyapunov time was about 10 million
years.  \citeN{Guzzo05} went on to corroborate the three-body
resonance theory by performing a large suite of integrations,
numerically detecting a large web of three-body resonances in the outer Solar System.

\citeN{MurrayHolman99} noted that the widths $\Delta a/a$ of the individual
resonant zones was of order $3\times 10^{-6}$, so that changes in the
ICs of that order can lead to regular motion.  They
note, however, that ``the uncertainties in the ICs,
and those introduced by our numerical model, are comfortably smaller
than the width of the individual resonances, so [the outer] Solar
System is almost certainly chaotic.''  Given that \citeN{Guzzo05}
has also detected many three-body resonances consistent with
Murry + Holman's theory, it would seem at first glance that chaos
in the outer Solar System is a fact.

However, the conclusion that the isolated outer Solar System is chaotic
cannot be taken for granted.  For example, it is known that symplectic
integration with too-large a timestep can inject chaos into an integrable
system \citep{HerbstAblowitz89,NewmanEtAlDDA2000}.  Although most authors verify their
primary results by performing ``checking'' integrations with smaller
timesteps, the checking integrations are not always performed for the
full duration of the main integrations.  This, combined with the fact
that longer symplectic integrations require {\em shorter} timesteps
\citep{BenettinGiorgilli94,Reich99} means that one cannot assume that a
timestep good enough, for example, for a 100-million-year integration
is also good enough for a 200-million-year integration.  There is
currently no known method for analytically choosing a short-enough
timestep {\it a priori}, and so the {\em only} method of verifying the
reliability of an integration is to re-perform the entire integration
with shorter-and-shorter timesteps until the results converge.
\citeN{NewmanEtAlDDA2000} used this method to demonstrate that, for a given
set of ICs, the \citeN{WisdomHolman91} symplectic mapping
with a 400 day timestep (about 11 steps per orbit, a commonly-quoted timestep)
admits chaos,
but that the results converge to regularity for any timestep less than
about 100 days.  However, many authors who find chaos have also
performed reasonable convergence tests, demonstrating that the
chaos does not always disappear at convergence.

There exists compelling evidence for the absense of chaos in the outer Solar System.
Laskar and others noted that when the entire solar system is integrated,
the inner Solar System manifests chaos on a 5-million-year timescale, but
the outer Solar System appears regular in these integrations.
Although Laskar's approximate theory can overlook some causes of chaos,
there also exist full-scale integrations that indicate the absense of chaos.
Grazier \etal (1999) and Newman \etal (2000) 
utilized a St\"{o}rmer integrator whose per-step numerical errors were bounded by the
double-precision machine epsilon (about $2\times 10^{-16}$), as long as the orbital
eccentricity is less than 0.5 and 1000 steps or more are taken per orbit.
Furthermore, their integration method took care to ensure that the roundoff
error was unbiased.
Note that, except for the possibility of having the same property with a
larger timestep, this is practically as good an integration as is possible using
double-precision.  Furthermore, such an integration is symplectic by default,
since it is essentially {\em exact} in double precision.  Using this
method, they performed an integration of the Jovian planets lasting
over 800 million years, and found no chaos.
\citeN{VaradiRunnegarGhil03} performed a 207My integration of the entire
Solar System, including even the effects of the Moon, and placed a lower
bound of 30My on the Lyapunov time of the system of Jovian planets.

We are thus left with the disturbing fact that, utilizing ``best
practices'' of numerical integration, some investigators integrate the
system of Jovian planets and find chaos, while others do not.

In this paper, we demonstrate that this apparent dilemma has a simple
solution.  Namely, that the boundary, in phase space, between chaos and
near-integrability is finer than previously recognized.  In particular,
the current observational uncertainty in the positions of the outer
planets is a few parts in $10^7$ \citep{Standish98-DE405,MorrisonEvans98}.
Within that
observational uncertainty, we find that some ICs lead to
chaos while others do not.  So, for example, drawing 7-digit ICs
from the same ephemeris at different times,
one finds some solutions that are chaotic, and some that are not.
Thus, different researchers who draw their initial coniditions from
the same ephemeris at different times can find vastly different
Lyapunov timescales.

\section{Methods}

With the exception of the two sets of initial conditions (ICs) we have
received from other authors \citep{MurrayHolman99,GrazierNewmanHymanSharp05} and the
set included in {\it Mercury 6.2} \citep{ChambersMercury99},
all ICs used in this paper are drawn at various epochs from
DE405 \citep{Standish98-DE405}, which is the latest planetary ephemeris
publicly available from JPL.  It has stated uncertainty for the positions
and masses of the outer planets of a few parts in $10^7$.
To ensure that our integration agrees over the short term with DE405,
we verified in several cases that we can integrate between different
sets of DE405 ICs, separated by as much as 100 years, while maintaining
at least 7 digits of agreement with DE405.

We integrate the system of Jovian planets using only Newtonian
gravity.  The inner planets are accounted for by adding their masses
to the Sun and perturbing the Sun's position and linear momentum to
equal that of the Sun+Mercucy+Venus+Earth+Mars system.  This ensures
that the resonances between the outer planets is shifted by an amount
that is second order in this mass ratio, roughly $3\times 10^{-11}$
\citep{MurrayHolman99}, which is far smaller than the uncertainty in the
outer-planet positions.  We assume constant masses for all objects and
ignore many effects which are probably relevant over a 200My timescale
(see for example \citeN{Laskar99}).
We account for solar mass loss at a rate of ${\dot m}/m\approx 10^{-7}$
per million years \citep{Laskar99,Noerdlinger05}, but note that we
observe no noticable difference if we keep the solar mass constant.

To reduce the possibility that our results are dependent on the
integration scheme,
we used three different numerical integration methods to verify many
of our results in this paper.  First, we used the Mercury 6.2 package
\citep{ChambersMercury99}, with the Wisdom-Holman \citep{WisdomHolman91}
symplectic mapping option (called {\tt MVS} in the input files).
We used stepsizes varying from 2 days to
400 days.  Second, we used the {\tt NBI} package,
which contains a 14th-order Cowell-St\"{o}rmer method with modifications
by the UCLA research group led by William Newman
\citep{GrazierNewman95,GrazierNewman96,VaradiRunnegarGhil03}.\footnote{
{\tt NBI} is available at {\tt http://astrobiology.ucla.edu/\~{}varadi/NBI/NBI.html},
or by searching the web for ``{\tt NBI Varadi}''.}
{\tt NBI} has been shown to have relative truncation errors below the
double precision machine epsilon (about $2\times 10^{-16}$) when more than
1000 steps per orbit are used and the orbital eccentricity is less than 0.5.
More precisely, if the largest component of the phase-space vector
of the solution at time $t$ has absolute value $M$, then the local errors
per step of each of the components are all less than $2\times 10^{-16}M$.
Note that this means that the component-by-component relative error
can be significantly greater than the machine precision for components
of the solution that are significantly less than $M$, but all components have
errors relative to $M$ which are smaller than the machine precision.
Furthermore, the authors of NBI have gone through great pains to ensure
that the roundoff error is unbiased.
We used a 4-day timestep for all {\tt NBI} integrations, which
gives more than 1000 timesteps per Jupiter orbit.
We have verified the
above ``exact to double precision'' property by comparison against
quadruple precision integrations described below.
Note that for the above-defined timestep, {\tt NBI} gives practically the best integration possible
using only double precision, and that such an integration is
symplectic by default since it essentially provides a solution which
is exact in double precision.  In our 200 million year integrations,
{\tt NBI} always had relative energy errors and angular momentum errors
of less than $2\times 10^{-11}$, with an average of about
$2\times 10^{-12}$.

Our third integrator was the {\it Taylor 1.4} package \citep{JorbaZou05}.
{\it Taylor 1.4} is a recent and impressive advance in integration
technology.  It is a general-purpose, off-the-shelf integrator which
utilizes automatic differentiation to compute arbitrary order Taylor
series expansions of the right-hand-side of the ODE.  {\it Taylor 1.4}
automatically adjusts the order and stepsize at each integration step
in an effort to minimize truncation error, and utilizes Horner's rule
in the evaluation of the Taylor series to minimize roundoff error.
As the authors note, integration accuracy is gained more efficiently
by increasing the order of the integration than by decreasing the
timestep, since the accuracy increases exponentially with the order
but only polynomially in the timestep.  Although {\it Taylor 1.4}
allows the user to specify a constant order and timestep, we chose to
allow it to use variable order and timestep while providing it with
a requested relative error tolerance equal to 1/1000 of the machine
precision, in order to produce solutions which were exact to within
roundoff error.  We found that {\it Taylor 1.4} typically used about 27th
order with about a 220 day timestep.  
The fact that the solution is exact to machine precision over such a long
timestep guarantees that accumulated roundoff is by far the smallest in
the {\em Taylor 1.4} integrations.  Furthermore, 
Taylor integrators are extremely stable when applied to non-stiff problems,
with the radius of convergence increasing linearly with integration order,
and in our case the timestep is well within the radius of convergence \citep{BarrioBlesaLara05}.
Finally, {\it Taylor 1.4} allows
the user to specify the machine arithmetic to use, including software
arithmetics.  Out-of-the-box, {\it Taylor 1.4} supports the use of IEEE
754 double precision (64 bit representation with a 53 bit mantissa),
Intel extended precision (80 bit representation with a 64-bit mantissa,
giving a machine precision of about $10^{-19}$, accessible as
{\tt long double} when using GCC on an Intel machine), the {\tt DoubleDouble}
datatype \citep{BriggsDoubleDouble} which provides software quadruple
precision in C++, and the GNU Multiple Precision Library, which
allows arbitrary precision floating point numbers in C++.  Most of our
integations using {\it Taylor 1.4} used Intel extended precision, which
is almost as fast as double precision and gives about 19 decimal digits
of accuracy.  Over our 200-million-year integrations using Intel extended
precision, Taylor typically had relative energy errors of less than
$8\times 10^{-14}$; the worst relative energy error observed in any of our
integrations was $2\times 10^{-13}$.
Integrations began with the Solar System's barycentre at the origin with zero velocity.
After 200 million years the barycentre drifted a maximum of
$3\times 10^{-10}$ AU,
while the $z$ component of the angular momentum was always conserved to
a relative accuracy better than $3\times 10^{-14}$.
We also performed a suite of quadruple precision integrations, in
which energy and angular momentum were each conserved to at least 26 significant
digits over 200 million years.
\clearpage
\begin{figure}[hbt]
\centering
\includegraphics[scale=0.5,angle=270]{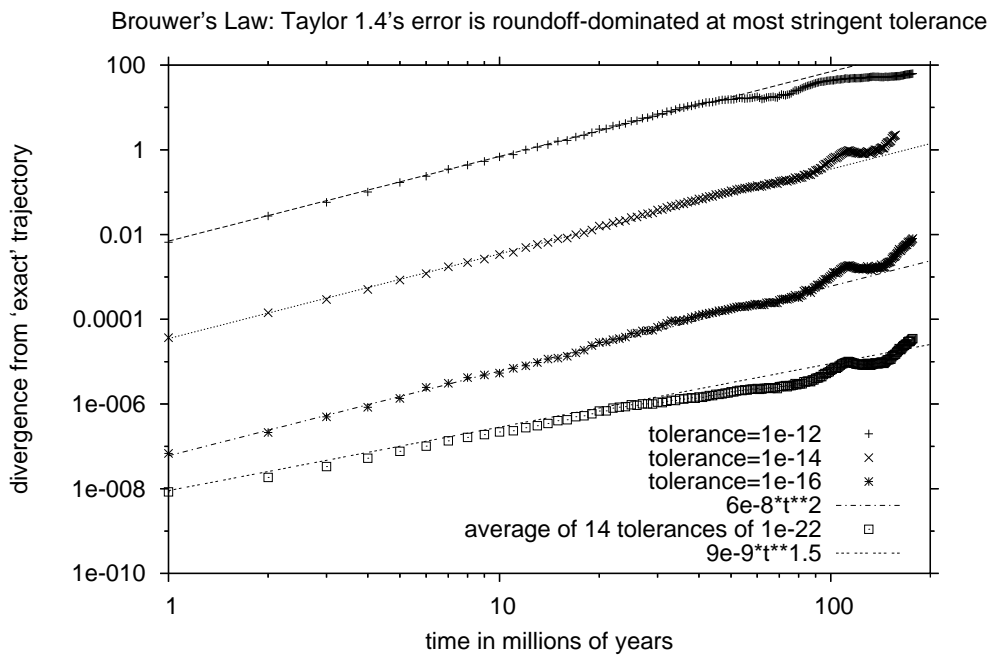}
\caption{{\it Taylor 1.4} satisfies Brouwer's Law:\ when local tolerance is set
above machine precision of $10^{-19}$,
phase-space error grows as $t^2$ due to biased truncation errors (upper three curves).
But with tolerance set well below the machine precision, its result is exact (\ie correctly
rounded) to machine precision, so that phase-space error grows only as $t^{1.5}$ (lowest
curve).}
\label{fig:TaylorBrouwer}
\end{figure}
\clearpage
Now we analyze the error growth as a function of time for our
non-symplectic integrators, applied to a non-chaotic IC.
One consequence of having a solution whose numerical error is dominated
by unbiased roundoff (ie., exact to machine precision) is that when
integrating a non-chaotic system, the total phase-space error grows
polynomially as $t^{1.5}$. If the local error is biased (either
by biased roundoff, or due to truncation errors in the integration
scheme), then the total phase-space error grows as $t^2$.  (This
is assuming that the integrator is not inherently symplectic, as is
the case with both NBI and {\it Taylor 1.4}.) This is
known as {\it Brouwer's Law} \citep{Brouwer37}.  As noted above,
\citeN{GrazierNewmanHymanSharp05} have demonstrated that the error in
{\tt NBI}'s integration is dominated by unbiased roundoff when using
1000 timesteps per orbit.  We tested {\it Taylor 1.4} in Intel extended
precision for similar properties by integrating the system of Jovian planets
for 200 million years using various integration tolerances up to and beyond
the machine precision of $10^{-19}$, and compared these integrations to
a {\it Taylor 1.4} integration that used quadruple precision.  In Figure
\ref{fig:TaylorBrouwer}, we plot the phase-space separation between
the quadruple precision integration and several integrations using
Intel extended precision.  We see that when the relative integration
tolerance is set above the machine precision, the error grows as
$t^2$ and is therefore truncation dominated.  But when the tolerance is
set to $10^{-22}$ (about a factor of 1000 below the machine precision), the
error grows as $t^{1.5}$, and is therefore dominated by unbiased roundoff.
This is consistent with {\it Taylor 1.4} producing results that are exact in
Intel extended precision when given a local relative error tolerance of
$10^{-22}$, just as {\tt NBI} produces exact results in double precision
when used with 1000 or more timesteps per orbit.  We see that after 200
million years, the errors in the positions of the planets are of order
$10^{-5}$ AU (in the case of non-chaotic ICs).  This translates into a phase
error, or equivalently an observational error, of substantially less
than one arc-minute.  A comparison of our Figure \ref{fig:TaylorBrouwer}
with Figure 2 of \citeN{GrazierNewmanKaulaHyman99} at their 200 million
year mark also demonstrates that {\it Taylor 1.4} using Intel extended
precision provides about 3 extra digits of precision over {\tt NBI}
using double precision, as would be expected when comparing 19-digit
and 16-digit integrations.  As we show later (Figure \ref{fig:final-ae-diffs}),
the differences in orbital elements are even smaller.

For comparison, we briefly mention the run-times of the integration
algorithms.  All timings are for a 2.8Ghz Pentium 4 processor.  As we
shall see later, the largest timestep for which the {\it Mercury 6.2}
integrations agreed with the others was 8 days; a 16-day timestep was
almost as good.  Thus we compare the Wisdom-Holman mapping with timesteps
of 16 and 8 days to NBI with a 4-day timestep, and {\it Taylor-1.4}.
In addition to the {\it Taylor-1.4} extended precision ({\tt long double})
timings used in this paper, we present its timings for standard IEEE
754 {\tt double} precision, for comparison to the other {\tt double}
precision integrations.  Table \ref{tab:runtime} presents the results.
We found that the total runtime was linearly proportional to the inverse
of the timestep, as would be expected.  We note several observations.
First, {\it Taylor-1.4} is not competitive in terms of efficiency.
However, note that {\it Taylor-1.4} is a ``proof-of-concept'' software
package for general-purpose integration of {\em any} system of ODEs,
and it currently generates code that can be quite inefficient.
A carefully hand-coded Taylor series integrator for Solar
System integrations is far more efficient, and can be competitive with
the above codes (Carles Simo 2007, personal communication).
Second, Wisdom-Holman with an 8-day timestep is the fastest case among
the integrations we tested that showed complete convergence.  Third,
Wisdom-Holman with a 4-day timestep (51 hours, not shown in the table)
is slower than NBI with a 4-day timestep.  Finally, although we did {\em
not} test NBI for convergence at timesteps less stringent than 4 days,
it is possible that NBI maintains the lead in being more efficient than
Wisdom-Holman, if it also shows convergence at larger timesteps.
\clearpage
\begin{table}
\begin{tabular}{|c|ccccc|}
\hline
integrator & WH dt=16d & WH dt=8d & NBI & Taylor-1.4 {\tt double} & Taylor-1.4 {\tt long double} \\
\hline
time (hours) & 13 & 26 & 40 & 100 & 150 \\
\hline
\end{tabular}
\caption{Run times for integrating the 5-body system for 200 million years on a 2.8Ghz pentium 4,
for {\it Mercury 6.2} with timesteps of 16 and 8 days, for NBI, and for {\it Taylor-1.4} in
{\tt double} and {\tt long double} precision.
}
\label{tab:runtime}
\end{table}
\clearpage
We will not directly measure the Lyapunov time in this paper, since it
is difficult to create an objective measure of the Lyapunov time over a
finite time interval.  Formally, the Lyapunov time is only defined over
an infinite time interval.  In practical terms, the divergence is almost
always polynomial for some nontrivial duration before the exponential
divergence emerges, and it is difficult to pinpoint the change-over
objectively.  However, when plotting the distance between two nearby
trajectories as a function of time, it is usually evident by visual inspection whether or
not exponential divergence has occured by the end of the simulation.
Thus we will plot the actual divergence between two numerical trajectories
initially differing by perturbing the position of Uranus by $10^{-14}$ AU
(about 1.5mm) in the $z$ direction.  We will call these pairs
of trajectories ``siblings.''  In the cases where we see only polynomial
divergence between siblings over a 200 million year integration, we will
abuse terminology and call these systems ``regular'', ``near-integrable'',
or ``non-chaotic'', although formally all we have shown is that the
Lyapunov time is longer than can be detected in a 200-million-year
integration.
\clearpage
\begin{figure}[p]
\centering
\includegraphics[scale=0.5,angle=270]{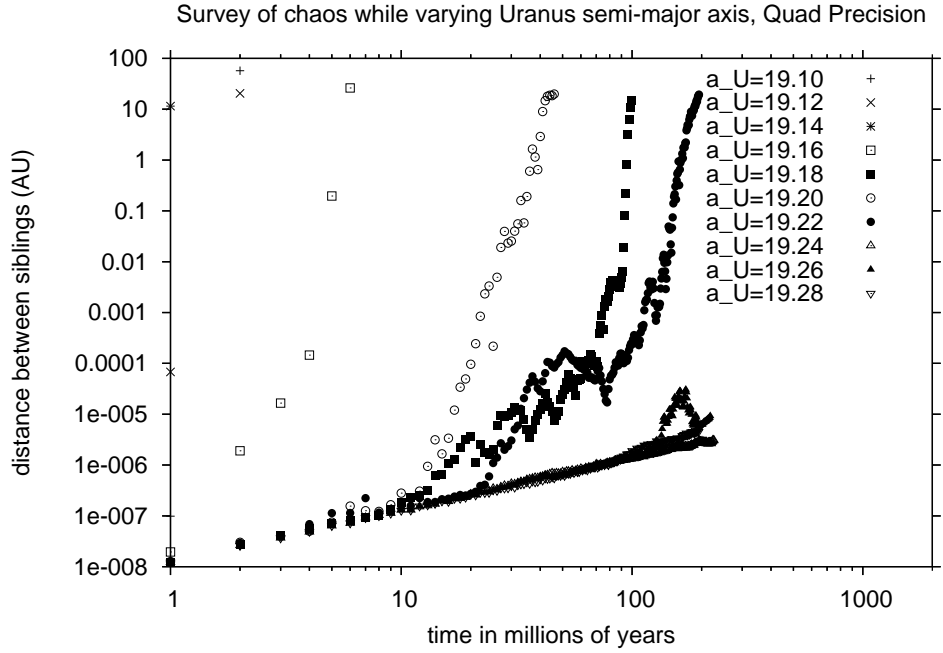}
\caption{Murray + Holman (1999) in quadruple precision.}
\label{fig:MH-dd}
\end{figure}
\begin{figure}
\centering
\includegraphics[scale=0.5,angle=270]{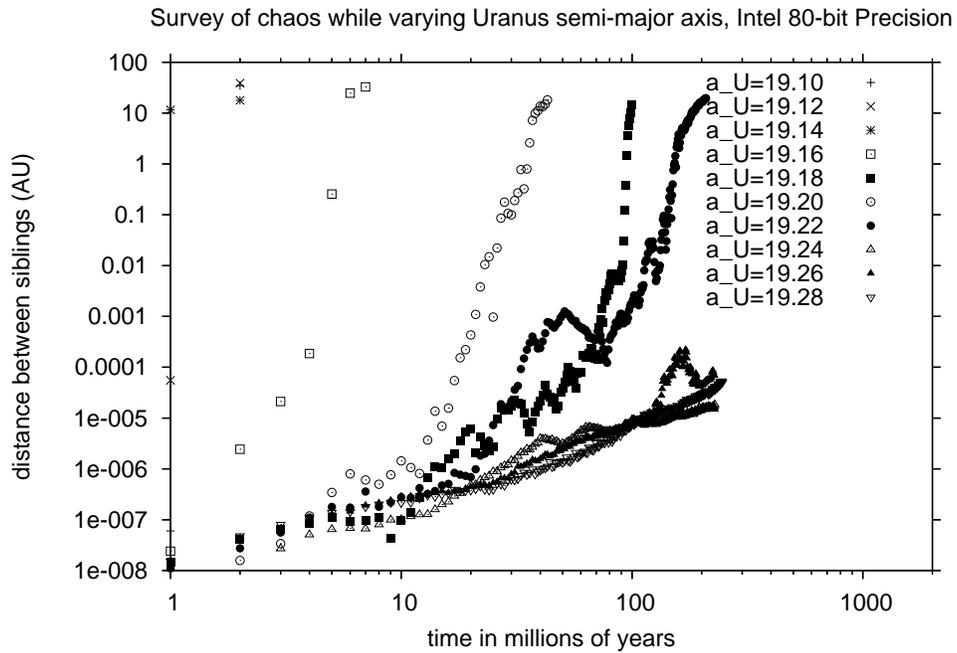}
\caption{Murray + Holman (1999) in {\tt long double}.}
\label{fig:MH-ld}
\end{figure}
\clearpage
As a prelude to our main results, we first crudely reproduce the results
of \citeN{MurrayHolman99}.  Murray + Holman performed a survey in which
the semi-major axis of Uranus $a_U$ was varied around its actual value,
plotting the measured Lyapunov time as a function of $a_U$.  Broadly, they
found that below $a_U\approx 19.18$, the Lyapunov time was substantially
less than $10^7$ years, sometimes as small as $10^5$ years; in the range
$a_U\approx 19.20--19.22$, the Lyapunov time fluctuated between about
1--10 million years, interspersed with cases in which the motion was
regular; near $a_U=19.24$ it was uniformly regular; near $a_U=19.26$
there was a tightly-packed region of both chaotic and regular orbits;
and near $a_U=19.28$ the motion was again uniformly regular.  We have
reproduced this survey using quadruple precision integrations,
using a very course grid in $a_U$ due to computing constraints.
(On a 2.8GHz Intel Pentium 4, a quadruple precision integration using
{\it Taylor 1.4} proceeds roughly at 100My per month of
CPU time.)
Figure \ref{fig:MH-dd} displays the distance between siblings,
for various values of $a_U$.  Although we
were not able to perform the survey using a finer grid in $a_U$ due
to the computational expense, we see that in broad outline we obtain
results similar to Murray + Holman.  In Figure \ref{fig:MH-ld},
we repeat the same survey using Intel extended precision.  We see that
the results are qualitatively identical, demonstrating that 19 digits of
precision (which requires about 1/20 of the CPU time of quadruple precision)
gives qualitatively identical, and quantitatively very similar, results.
Henceforth in this paper, all {\it Taylor 1.4} integrations are
performed using Intel extended precision, with the relative local error
tolerance set to $10^{-22}$.

\section{Results}
\label{sec:results}

\subsection{Corroborating previous results}

As an early step, the author obtained from Murray + Holman
their initial conditions (ICs) used in \citeN{MurrayHolman99}, and
verified using accurate integrations that their system was chaotic.
The author then obtained from P. Sharp the ICs used by
\citeN{GrazierNewmanHymanSharp05}, and verified that their system was regular,
at least over a 200My timespan.  The author then spent significant
time eliminating several possible reasons for the discrepancy, such as
incorrect ICs, incorrect deletion of the inner planets,
\etc  The author also integrated from one set of ICs
to the other (having along the way to account for the fact that they were
in different co-ordinate systems), and finally found that the systems agreed
with each other to a few parts in $10^7$ when integrated to the same epoch.
After some time it became apparent that neither group of authors had made
any obvious errors.  The conclusion seemed to be that both systems
nominally represented the outer Solar System to within observational
error, but one was chaotic and one was not.

\subsection{Ephemeris Initial Conditions Drawn at Different Times}

It is reasonable to make the assumption that either our Solar System is
chaotic, or it is not.  It cannot be both.  This appears to be the assumption
that most practitioners make when measuring ``the'' Lyapunov time of
our Solar System.  Although this is probably
a reasonable assumption, it does not follow that all initial conditions (ICs)
drawn from an ephemeris are equivalent.  The most recent ephemeris published
by JPL is DE405 \citep{Standish98-DE405}.  It is based upon hundreds of thousands
of observations, all of which have finite error.  Thus, the ephemeris does
not represent the exact Solar System.  For the outer planets, the best match
of DE405 to the observations yields residual errors of a few parts in $10^7$
\citep{Standish98-DE405,MorrisonEvans98}.
The masses of the planets are also known to only a few parts in $10^7$.
It is possible that within these error bounds there exist different solutions
with different Lyapunov times.  In particular, it may be possible that some
solutions display chaos on a 200-million-year timescale, while others do not.
\clearpage
\begin{figure}[p]
\centering
\includegraphics[totalheight=7.5in]{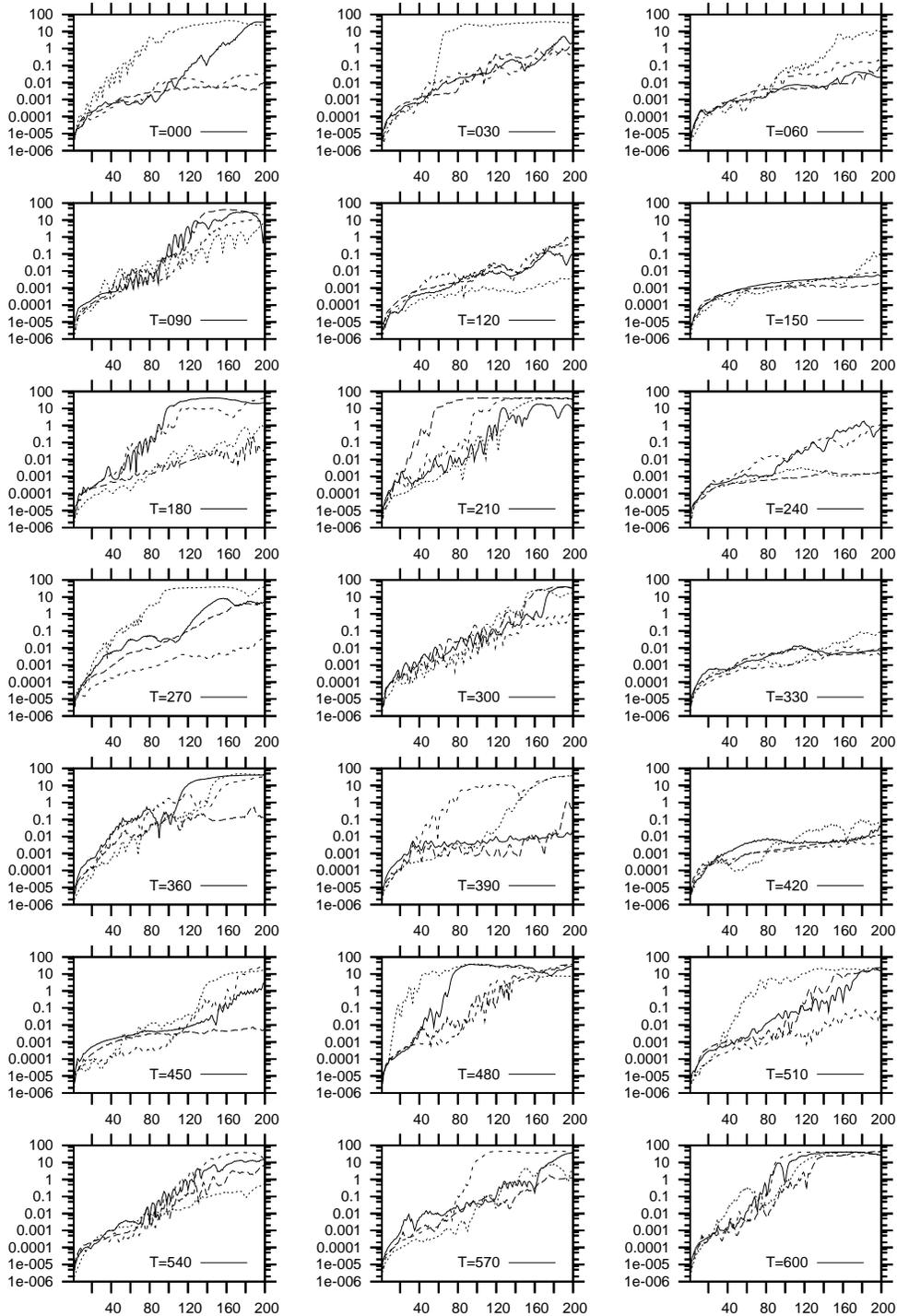}
\caption{\small Distance in AU between ``sibling'' trajectories as a function
of time for 21 sets of initial conditions drawn
from DE405 at 30-day intervals.  Each plot represents one initial
condition.  Siblings are integrated with the Wisdom-Holman symplectic map
with timesteps: 50 (solid line), 100 (long dash), 200 (short dash),
and 400 days (dotted).  For any given plot, note disagreement across timesteps,
and that the switch between chaos and regularity is not always monotonic with timestep.}
\label{fig:multiplot-big-steps}
\end{figure}
\clearpage
To test this hypothesis, we drew 21 sets of ICs from DE405,
at 30-day intervals starting at Julian date 2448235 (9.5 Dec, 1990).
We represent each of our 21 sets of ICs using a 3-digit numeral, 000,
030, 060, $\ldots$, 570, 600, representing the
number of days after JD2448235 at which the intitial condition is drawn.
We chose a 2-year total interval across which to draw our ICs in order
to to ensure a reasonable sample of inner-planet positions before deleting
the inner planets (a Martian year is about 2 Earth years).
We drew the initial conditions by taking the output of the program
{\tt testeph.f}, which is included with the DE405 ephemeris.  The output
of {\tt testeph.f} is rounded to 7 digits, since no more digits are
justifiable.\footnote{Although this is probably not the
best way to uniformly sample initial conditions from within the error ball
representing the error in the observations, it {\em does} represent a
reasonable way to reproduce how users of DE405, taking initial conditions
from DE405 to 7 digits, get their samples.} As
described above, we augment the mass of the Sun with that of the inner planets
and augment the Sun's position and linear momentum to match that of the inner
Solar System.  For each of the 21 ICs, we generate a ``sibling''
IC which is offset by $10^{-14}$ AU (1.5mm).  We then
integrate both of them for 200 million years, and measure the distance between
them at 1-million-year intervals.  To ensure that our results are not integrator-
or timestep-dependent, we perform each integration in several ways:
first, with the Wisdom-Holman symplectic mapping included with {\it Mercury 6.2}
\citep{ChambersMercury99} with timesteps of 400, 200, 100, 50, 32, 16, 8, 4,
and 2 days; second, with {\tt NBI} with a timestep of 4 days; and third,
with {\it Taylor 1.4} in Intel extended precision with a relative error
tolerance of $10^{-22}$.
Results for the Wisdom-Holman integrations with larger timesteps are displayed in
Figure \ref{fig:multiplot-big-steps}.  After staring at these figures for
awhile, several key observations become apparent.  First, when comparing
across the 21 sets of initial conditions, there is
remarkable disagreement about whether or not the outer Solar System displays exponential
divergence.  This will be discussed further below, using more accurate integrations.
Looking at an individual graph, but across timesteps, we note that there are very few cases
(e.g. $t=150,330,420$)
in which there is universal agreement between all four timesteps
that divergence is polynomial.  There are also only a few cases that
universally agree that the divergence is exponential.
In most cases, there is substantial disagreement across timesteps whether a particular
initial condition admits chaos.  This is consistent with the observation of
Newman \etal (2000).  However, in contrast to Newman \etal (2000), we
note that the ``switch'' from chaos to non-chaos is not always monotonic
in timestep.
For example, for system {\tt 000}, timesteps of 400 and 50 days admit chaos, while
the ``in-between'' timesteps of 200 and 100 days do not.  For system {\tt 180},
timesteps of 400 and 200 days display non-chaos, while for 100 and 50 day timesteps,
chaos is apparent; this is precisely opposite to what one would expect if
large timesteps were injecting chaos into the system.  As we shall see later
in the paper (Table \ref{tab:chaosCounts}), there appears also to be no
observable correllation between
the timestep and the percentage of ICs that admit chaos.  Thus we hypothesize
that the discrepancy across timesteps is due more to perturbations in
the ICs and our use of only low-order symplectic correctors
\citep{WisdomHolmanTouma96,ChambersMercury99},
than it is due to unreliable integration at large timesteps.  Corroborating
this hypothesis would require us to re-perform these experiments using
higher-order symplectic correctors or ``warmup'' \citep{SahaTremaine92},
which is a possible direction for future research.
\clearpage
\begin{figure}[p]
\centering
\includegraphics[totalheight=7.5in]{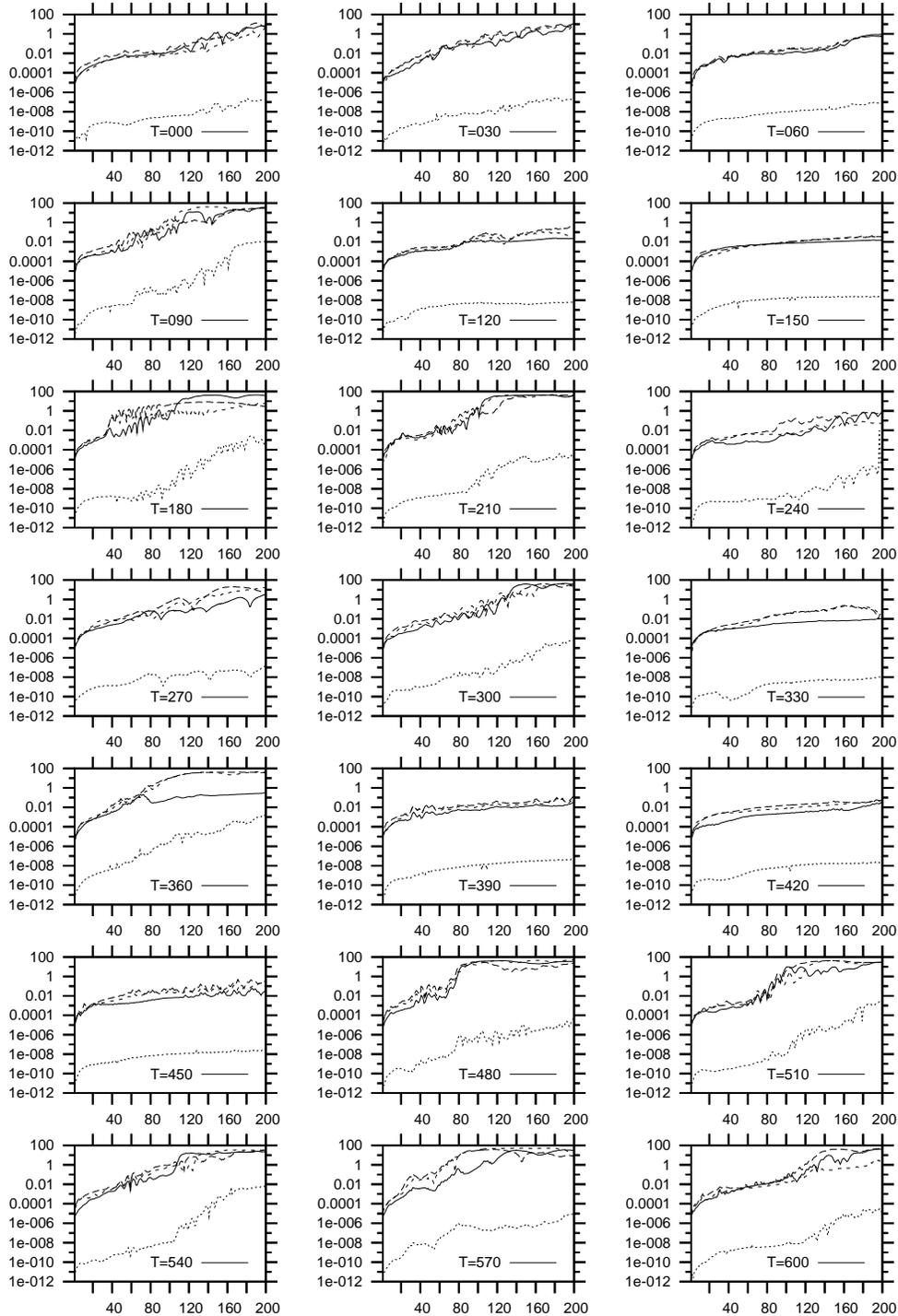}
\caption{\small Similar to previous Figure, but more accurate integrations.
{\tt NBI}: solid lines; Wisdom-Holman: dashed lines with $dt=$8 days (long dash),
4 days (short dash);
Dotted lines ({\it Taylor-1.4}) are shifted
below others because {\it Taylor-1.4} is more accurate.
For any given IC, there is good
agreement between the shapes of the curves, indicating agreement on
the existence (or lack) of chaos.  (Divergence saturates at $\approx$ 100 AU.)
}
\label{fig:multiplot-logy}
\end{figure}
\begin{figure}[p]
\centering
\includegraphics[totalheight=7.5in]{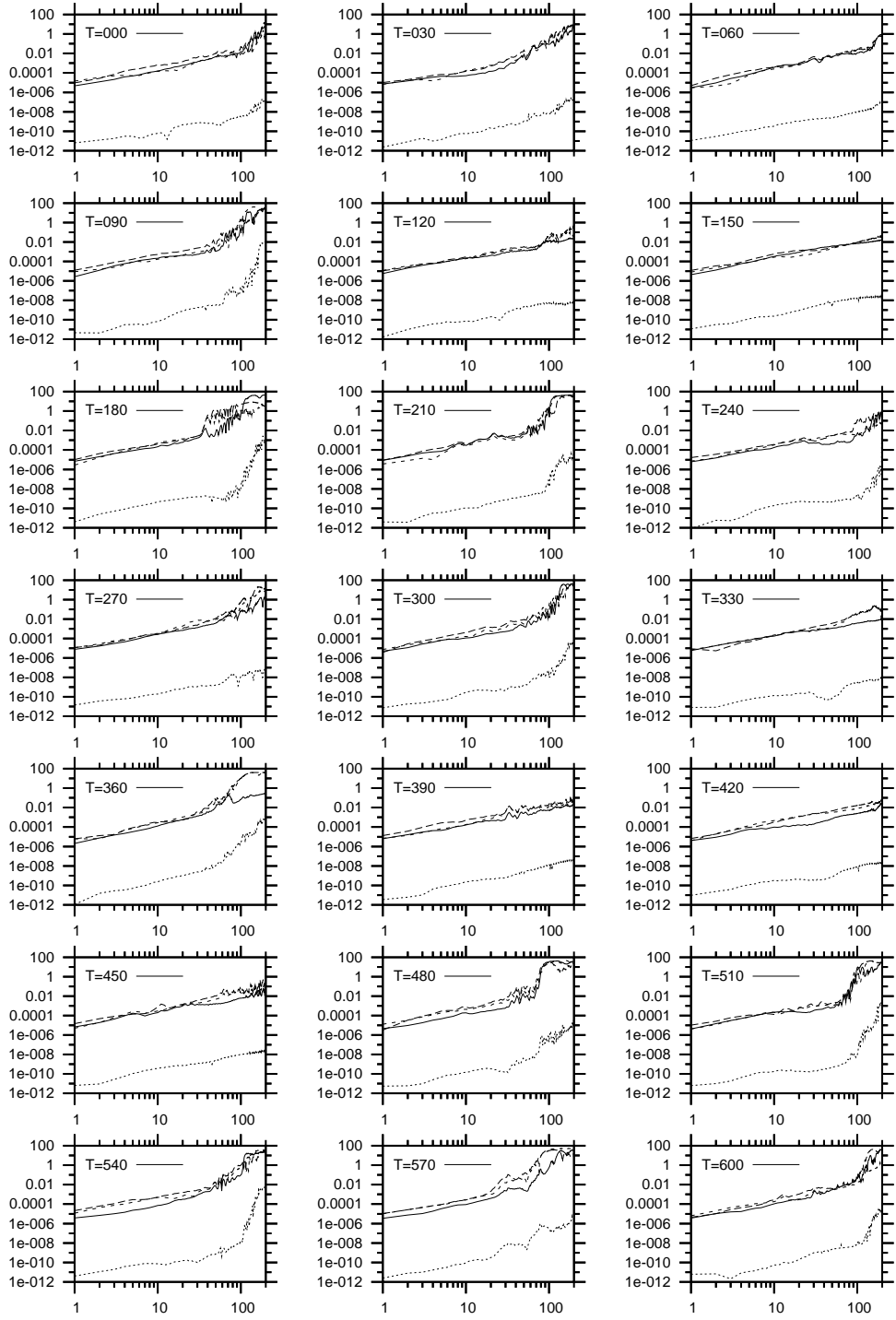}
\caption{Curves identical to the previous figure, but plotted on a log-log scale,
which depicts polynomial divergence as a straight line.}
\label{fig:multiplot-log}
\end{figure}
\clearpage
Figures \ref{fig:multiplot-logy} and \ref{fig:multiplot-log} plot the
divergence between sibling trajectories for all 21 ICs, as integrated
by more accurate integrations showing convergence: {\tt NBI},
the Wisdom-Holman mappings with timesteps of 8 and 4 days, and {\it
Taylor 1.4}.  Wisdom-Holman with 2 day timesteps agreed with these curves,
but are omitted to reduce clutter; Wisdom-Holman with a timestep of
16 days also showed good agreement in all but two cases.
Both figures are identical except that Figure
\ref{fig:multiplot-logy} uses a log-linear scale, while Figure \ref{fig:multiplot-log}
uses a log-log scale; different features are visible using the different scales.
After staring at these graphs for awhile, at least two observations present
themselves.  First, if one looks at the system corresponding to
any single IC, there is usually good agreement between the
integrations as to the future divergence between the sibling trajectories
of that particular case.  This demonstrates that convergence has occured
and makes it unlikely that the results are
integrator dependent.  Second, looking across cases, it appears that the
future of the outer Solar System over the next 200 million years is quite
uncertain, varying from nearly integrable, to chaotic with a Lyapunov time
of order 10 million years or less.  This is quite a startlingly diverse
array of possible outcomes, considering that the ICs
for these systems are all drawn from the same ephemeris, all less than
2 years apart, and presumably differing from each other by only a few
parts in $10^7$.  The author has, in fact, verified that several of
the ICs, when all integrated up to the same epoch in the
vicinity of 1991, agree with each other to a few parts in $10^7$.
\clearpage
\begin{figure}[hbt]
\centering
\includegraphics[scale=0.5,angle=270]{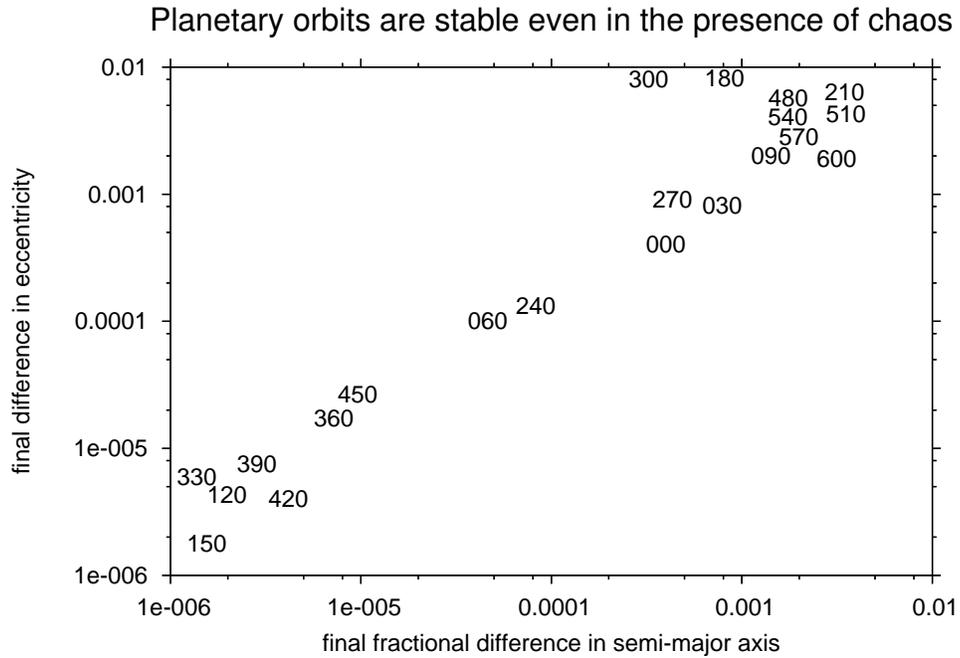}
\caption{
The final difference at the 200My mark between siblings,
in semi-major axis and eccentricity, for the 21 cases depicted in the
previous Figures.
The integrator was NBI.  Eccentricity difference
is the Euclidean distance between siblings in the 4-dimensional space
consisting of the four orbital eccentricities.  Semi-major axis difference
is the Euclidean distance between siblings in the 4-dimensional space
consisting of the four $\Delta a/a$ values.  Some values have been moved
slightly for textual clarity, but in no case by more than the width or
height of a character.}
\label{fig:final-ae-diffs}
\end{figure}
\clearpage
Figures \ref{fig:multiplot-logy} and \ref{fig:multiplot-log} demonstrate
that in the chaotic cases, the siblings can have their respective planets on
opposite sides of the Solar System after 200My.  However, Figure
\ref{fig:final-ae-diffs} demonstrates that changes in the orbital elements 
are much less drastic, demonstrating that the Solar System is practically
stable over a 200My timescale even when it is chaotic.

\subsection{Explicitly Perturbed Initial Conditions}

Following \citeN{MurrayHolman99}, we performed several surveys
in which we perturbed the semi-major axis $a_U$ of Uranus from its
current value, but kept all other initial conditions (ICs) constant.
We used all three previously mentioned integrators; Wisdom-Holman
with timesteps of 4 and 8 days; {\tt NBI}; and {\it Taylor-1.4}.
The IC was the default one from the file {\tt big.in}
included with {\it Mercury 6.2} \citep{ChambersMercury99}, which according
to the documentation is from JD2451000.5.  The inner planets were
deleted, with their mass and momentum augmenting the Sun's as described
elsewhere in this paper.
We completed surveys in which $a_U$ was changed
in steps of $2\times 10^{-k}$ for $k=6,7,8$, which corresponds
to $\Delta a_U/a_U$ in steps of $10^{-(k+1)}$.  We went 10 steps
in each direction for each value of $k$.  For each step, we
generated a ``sibling'' IC by randomly perturbing the positions
of {\em all} planets by an amount bounded by $10^{-14}$ AU.\footnote{
Note that, for no good reason, this is different from the perturbations
used to generate siblings in the rest of the paper.}
We then integrated both for 200My, and plotted the distance between them
as a function of time.  For $k=7,8$, there was no significant
difference between any of the integrations.  That is, all siblings
at all steps had virtually identical divergences when changing
$a_U$ in 10 steps of $2\times 10^{-\{7,8\}}$ AU, for a total change of
$2\times 10^{-6}$.  This corresponds to $\Delta a_U/a_U$ stepped by $10^{-\{8,9\}}$,
for a total change in $\Delta a_U/a_U$ of $10^{-7}$ in each direction.
However, for $k=6$, some of the steps showed chaos while others did not.
The change was not monotonic: over the 21 steps (10 in either direction plus
the ``baseline'' case), there were three ``switches'' between chaos and stability.
\clearpage
\begin{figure}[hbt]
\centering
\includegraphics[scale=1.3]{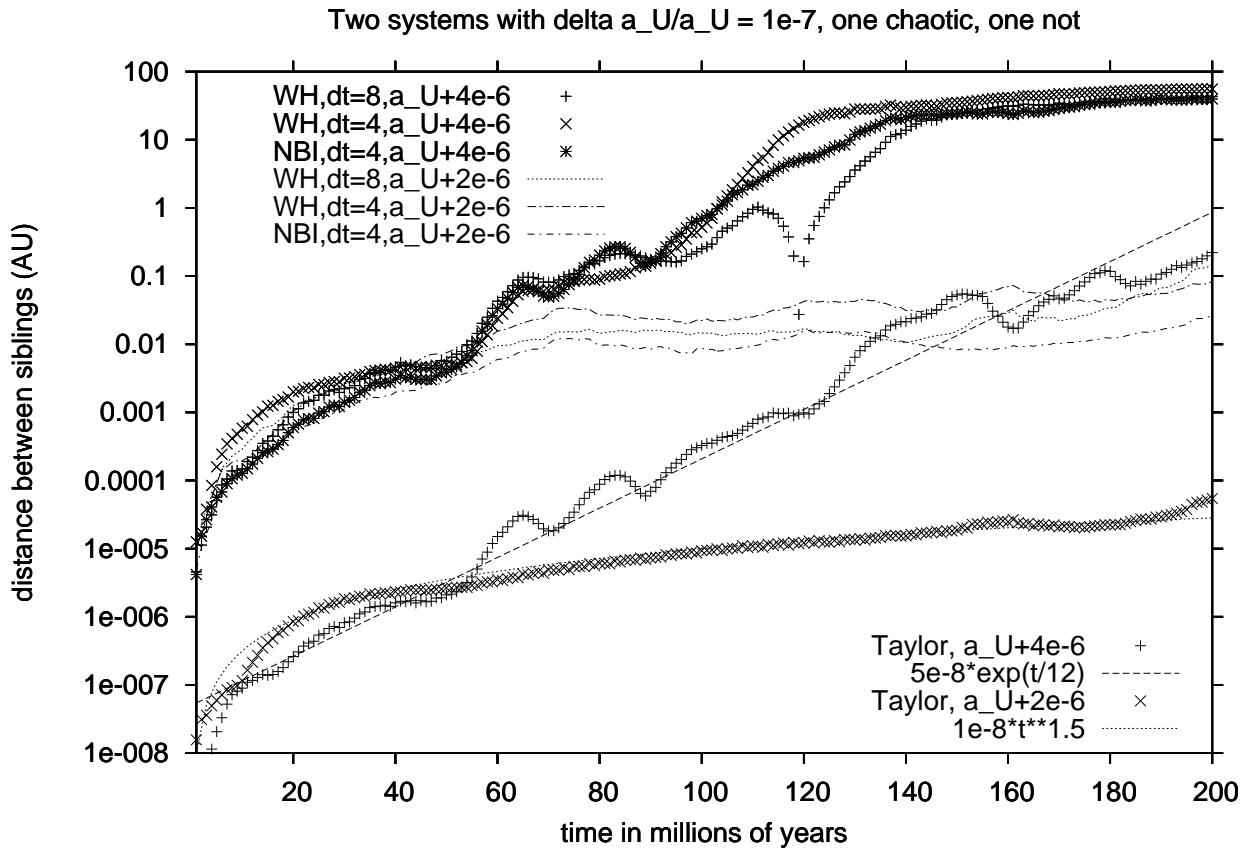}
\caption{Two systems, both using ICs within the error bounds of the best known positions
for the outer planets.  Both have identical initial conditions except for the
semi-major axis $a_U$ of Uranus, which differs between the two systems by
$10^{-7}$ in $\Delta a_U/a_U$.  One admits chaos, while the other does not.
The ``upper'' six curves (all starting with sibling distances near $10^{-5}$)
are all double-precision integrations, two using Wisdom-Holman with
timesteps of 4 and 8 days, and one using {\tt NBI}.
Of the six, the three plotted with points are the chaotic trajectory and
the three plotted with lines are the non-chaotic trajectory.
The ``lower'' two curves (starting with sibling distances near $10^{-8}$)
are integrated in extended precision with {\it Taylor 1.4}.
The chaotic one fits an exponential curve with a Lyapunov time of about 12 million
years, while the non-chaotic one has the two trajectories separating approximately
as $t^{1.5}$.}
\label{fig:surprise-all}
\end{figure}
\clearpage
Figure \ref{fig:surprise-all} plots two of these 21 systems.
The value of $\Delta a_U/a_U$ differs between the two systems by one part in $10^7$.
One of the systems appears chaotic, and the other does not,
over a 200My timespan.  The non-chaotic one has a semi-major
axis of $a_U + 2\times 10^{-6}$, while the chaotic one has semi-major
axis $a_U+4\times 10^{-6}$.  All other ICs in the two
systems are identical.  To ensure that the result is not integrator
dependent, we have repeated the integrations with the Wisdom-Holman
mapping included in {\it Mercury 6.2} with timesteps of 8 and 4 days;
and with {\tt NBI} with a timestep of 4 days.  As can be seen, all the
integrations agree quite well with one another.  Note that the
{\it Taylor 1.4} integrations provide 3 extra digits of precision,
and so the curves for {\it Taylor 1.4} are displaced about 3 orders
of magnitude below the curves computed in double precision.  Otherwise
the shapes of the curves are virtually identical.  We note that the
chaotic one has a Lyapunov time of about 12 million years, while the
regular one has the sibling trajectories separating from each other
polynomially in time as $t^{1.5}$.

\subsection{Accurate integrations over the age of the Solar System}

The author has reported related results for integrations lasting
$10^9$ years \citep{HayesNaturePhysics07}.  The essential conclusion
is the same, in that even after $10^9$ years, there remain some
ICs (about 10\%) that show no evidence of chaos, although {\em some}
of the ICs appearing as regular over 200My develop exponential
divergence later.

\begin{figure}[hbt]
\centering
\includegraphics[scale=1,angle=0]{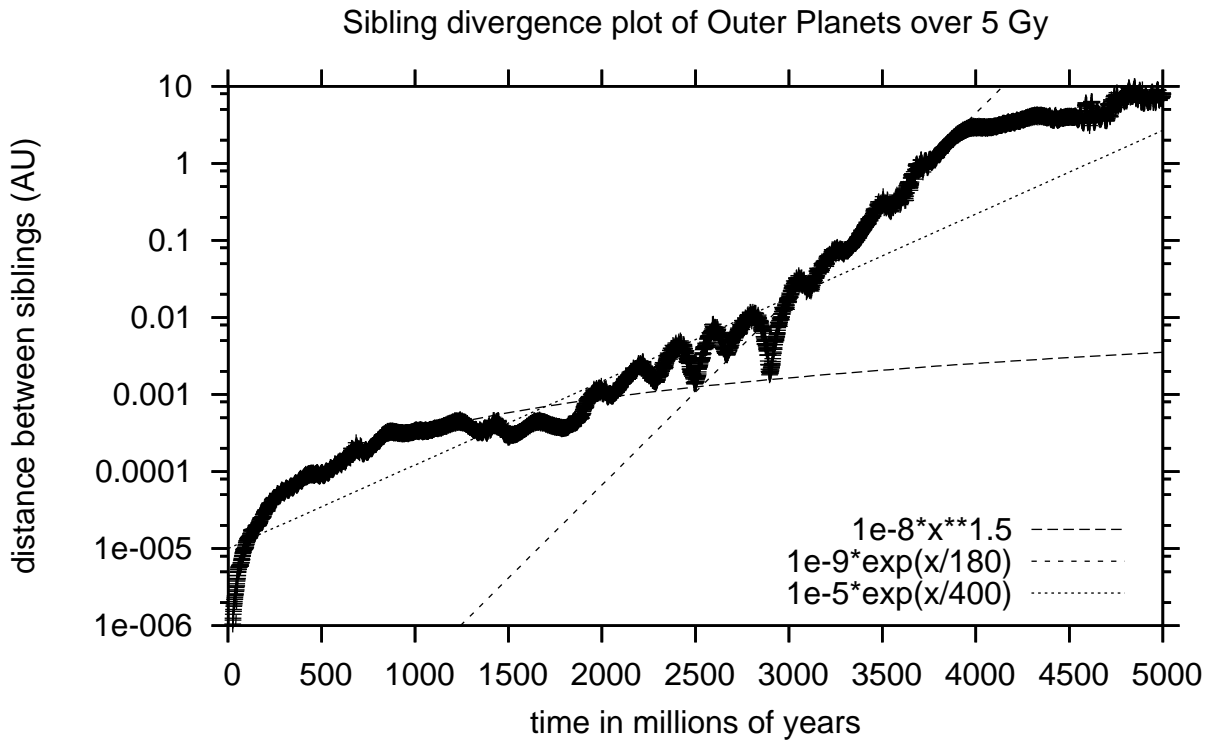}
\caption{Sibling divergence over 5 Gy of the canonical IC of DE405, integrated using Taylor-1.4.}
\label{fig:5Gy}
\end{figure}

Figure \ref{fig:5Gy} displays the sibling divergence over $5\times 10^9$
years of the ``canonical'' IC used by DE405 (JED 2440400.5, June 28,
1969).  As we can see, this IC shows little evidence of chaos for about
the first 1.5Gy, and then develops slow exponential divergence with a
Lyapunov time between about 200 and 400 million years.  The individual
planets each show similarly-shaped divergence curves (not shown), with the magnitude
of divergence increasing with orbital radius.  After 5 Gy, the uncertainty
in Jupiter's position for this IC is less than 1 AU, while the uncertainty
in Neptune's position is about 9 AU.  Thus, there is a non-negligible
chance that, if the Solar System lies close enough to the ``canonical''
IC of DE405, that we can know within about 10--15 degrees where each outer
planet will be in its orbit when the Sun ends its main-sequence lifetime
and becomes a red giant.
Note that the levelling-off that starts at about the 4 Gy mark is {\em
not} saturation (which occurs closer to 100 AU separation, while the separation
at 5 Gy here is less than 10);
the outer Solar System instead seems to be entering again into
a period of polynomial (non-chaotic) divergence.

\clearpage

\subsection{Percentage of initial conditions displaying chaos}

Observing the distance between sibling trajectories in Figures \ref{fig:multiplot-logy} and
\ref{fig:multiplot-log} at the 200My mark, we can reasonably simplify the distinction between
chaotic and regular trajectories.  By choosing a cutoff distance $C$
and restricting our view to the various double-precision integrations,
we can claim that siblings differing by less than $C$ after 200My
are regular (in the sense of having no observable Lyapunov exponent),
while those differing by more than $C$ are chaotic with a measurable
positive Lyapunov exponent.  Table \ref{tab:chaosCounts} lists the number
of systems that are chaotic by the above definition, as a function of
timestep and cutoff.  In addition to the 21-sample group of
ICs displayed in Figures \ref{fig:multiplot-logy} and \ref{fig:multiplot-log},
we also drew a second sample group of 10 samples, spaced at 10-year intervals
from 1900 to 1990.
Since the {\tt testeph.f} program included in DE405
provides only 7 digits of precision (corresponding to the accuracy
to which the positions are known), taking ICs from DE405 at different times
effectively takes ICs from different exact orbits, differing from
each other by as much as one part in $10^7$.
(As noted in a previous footnote, this method of sampling may not ideally represent
an unbiased sample from the observational error ball;
instead, it is an unbiased sample from the set of
7-digit-rounded initial conditions drawn from DE405.)
We make several observations.
First, the fraction of sampled systems that are chaotic by this
simple definition is roughly about ($70\pm 10)\%$, and is relatively
independent of both the timestep and the two sample groups (30-day \vs 10-year samples),
although it of course increases as we decrease the cutoff.
Recall that for a given initial condition, different stepsizes can give
different results, so that {\em which} systems are chaotic
changes as the timestep changes.  However, here we measure only the
{\em number} of chaotic systems as a function of timestep.
If chaos
were being ``injected'' into the system by the integrator, we would
expect that the number of systems displaying chaos should increase
with increasing timestep.  However, this is not observed. A
possible interpretation of this is that
even a 400 day timestep reliably determines whether {\em some} system is chaotic,
but the system is slightly different for each timestep due to the
IC perturbation introduced by the symplectic
integration \citep{SahaTremaine92,WisdomHolmanTouma96}.
Verifying this would require us to implement ``warmup'' \citep{SahaTremaine92} or
higher-order symplectic
correctors than are included with {\it Mercury 6.2} \citep{ChambersMercury99},
which we have not done.  However, the fact that the fraction
of systems displaying chaos is independent of timestep argues against the
``chaos is injected by the integrator'' hypothesis, at least for the timesteps
used in this paper.

\clearpage
\begin{table}
\begin{tabular}{c|cccc|cccc}
{} &	{} &	{C=1} &	{} & {} &	{} &{C=0.1}	 &	{} &	{} \\

dt &	/10 &	/21 & total	 & \%	 &	/10 &	/21 & total & \% \\
\hline
400 &	7 &	14 &	21 &	67.7 &	9 &	16 &	25 &	80.6 \\
200 &	7 &	11 &	18 &	58.1 &	8 &	15 &	23 &	74.2 \\
100 &	6 &	11 &	17 &	54.8 &	8 &	13 &	21 &	67.7 \\
050 &	6 &	14 &	20 &	64.5 &	8 &	15 &	23 &	74.2 \\
032 &	7 &	10 &	17 &	54.8 &	9 &	15 &	24 &	77.4 \\
016 &	7 &	14 &	21 &	67.7 &	8 &	16 &	24 &	77.4 \\
008 &	8 &	13 &	21 &	67.7 &	8 &	14 &	22 &	71.0 \\
004 &	8 &	13 &	21 &	67.7 &	8 &	18 &	26 &	83.9 \\
002 &	8 &	13 &	21 &	67.7 &	8 &	15 &	23 &	74.2 \\
\end{tabular}
\caption{The percentage of initial conditions drawn from DE405 that
admit chaos.  Column labels:
$C$ is the cutoff described in the text;
$dt$ is the timestep; /10 means
``out of the 10 initial conditions drawn at 10-year intervals from 1900
to 1990''; /21 means ``out of the 21 initial conditions drawn at 30-day
intervals starting at 1990''; {\it total} is the sum of the previous
two columns;
\% is the percentage of systems out of the 31 that display chaos according
to the cutoff.
}
\label{tab:chaosCounts}
\end{table}
\clearpage
\subsection{Chaos in the inner solar system is robust}

\citeN{VaradiRunnegarGhil03} performed
a 207My integration of the entire Solar System, including some
non-Newtonian effects and a highly-tuned
approximation to the effects of the Moon, and placed a lower bound of
30My on the Lyapunov time of the outer Solar System.
However, they still saw chaos in the inner solar system.
To test the robustness of chaos in the inner solar system, we performed
several integrations of 8 planets (Mercury through Neptune) using
the Wisdom-Holman mapping with timesteps of 8, 4, 2, 1, and 0.5 days.
We treated the Earth-Moon system as a single body.  To ensure that
chaos in the outer solar system did not ``infect'' the inner solar system,
we used only those DE405 ICs from Figures \ref{fig:multiplot-logy} and
\ref{fig:multiplot-log} for which the outer solar system was
regular over 200My.  We then
integrated the system until chaos appeared.  In all
cases, even though the outer solar system was regular, the inner
solar system displayed chaos over a short timescale such that information
about the inner planetary positions was lost within about 20--50My.
Thus, unlike the outer Solar System, we observe that chaos in the
inner solar system is robust.

\section{Discussion}

We conclude that perturbations in the initial conditions (ICs) of the
outer planets as small as one part in $10^7$ can change the behaviour
from regular to chaotic, and back, when measured over a timespan
of 200 million years.  We believe this is the first demonstration
of the ``switch'' from chaos to near-integrability with such a
small perturbation of the ICs.  Since our knowledge of the orbital
positions of the outer planets is comparable to one part in $10^7$,
it follows that, even if our simplistic physical model accounting only
for Newtonian gravity were the correct model, it would be impossible at
present to determine the Lyapunov time of the system of Jovian planets.
Furthermore, it implies that an IC with 7 digits of precision (which
is all an ephemeris can justifiably provide) can randomly lie on a
chaotic or non-chaotic trajectory.  Since our results converge in the
limit of small timestep for the Wisdom-Holman mapping, and the converged
results also agree with two very different high-accuracy integrations,
and finally since the high-accuracy integrations in turn agree very well
with quadruple-precision integrations, we believe that the results in
this paper are substantially free of significant numerical artifacts.

\citeN{Guzzo05} corroborates the existence of a large web of 3-body
resonances in the outer Solar System, and finds that their placement is
consistent with Murray and Holman's (1999) theory.  Guzzo used his own
fourth-order symplectic integrator \citep{Guzzo01}, and performed what
appear to be reasonable convergence tests to verify the robustness of
his main results.  Thus, Murray and Holman's theory appears to explain
the existence and placement of 3-planet resonances.  Furthermore, chaotic
regions in Hamiltonian systems are usually densely packed with both chaotic
and regular orbits \citep{LichtenbergLieberman92}.
This paper corroborates the observation of densely
packed regular and chaotic orbits, at a scale previously unexplored for
the system of Jovian planets.

As discussed in the text relating to Figure \ref{fig:surprise-all}, we
performed surveys across the semi-major axis $a_U$ of Uranus in steps of
$2\times 10^{-k}$ AU for $k=6,7,8$.  We found that, around the current
best-estimate value of $a_U$, perturbations smaller than $2\times 10^{-6}$
AU had no effect on the existence of chaos.  However, this does not imply
that perturbations this small cannot have an effect; it simply means
that the ``border'' between chaos and regularity is not within $2\times
10^{-6}$ AU of the current best estimate of $a_U$.  However, it is clear
that the border between chaos and regularity {\em is} between the two
systems depicted in Figure \ref{fig:surprise-all}, which differ from each
other in $a_U$ by $2\times 10^{-6}$ AU.  Although a survey across $a_U$
for values between those two systems may not be very relevant from a
physical standpoint, it might be very interesting from a dynamical systems
and chaos perspective to probe the structure of the border between chaos
and regularity.  In particular, it may be interesting to see if the
border itself has some sort of fractal structure \citep{MandelbrotBook82}.
Such chaotic structure has already been observed in the circular
restricted three-body problem \citep{Murison89}.

Newman \etal (2000) gave a compelling demonstration
that the Wisdom + Holman symplectic mapping with too-large a timestep
could introduce chaos into a near-integrable system
by first showing that they could reproduce the chaos with a 400-day timestep,
and then showing that the integration converged to being regular with
a timestep of about 50 days or less.  Our results appear to hint
that an even smaller timestep may be required: our curves depicting
divergence-of-nearby-orbits did not fully converge until the timestep was
8 days or less.  On the other hand, the symplectic integrators produce
solutions that effectively integrate a system with slightly perturbed ICs,
although these perturbations decrease with decreasing timestep
\citep{SahaTremaine92,WisdomHolmanTouma96}.
Our Figure \ref{fig:multiplot-big-steps} demonstrates
that the behaviour does not always ``switch'' monotonically in timestep
from chaotic to non-chaotic as observed by Newman \etal
(2000); Table \ref{tab:chaosCounts} also hints that the
``amount'' of chaos does not appear to increase with increasing timestep.
An alternate interpretation is that even a 400-day timestep accurately
integrates {\em an} orbit, but not {\it the} orbit that we chose.
In particular, it may accurately integrate an orbit whose IC is perturbed
slightly from the one we chose,
and an appropriate correction may allow us to recover the correct orbit using
an integration with much larger timestep \citep{SahaTremaine92,WisdomHolmanTouma96}.
However, the symplectic correctors in Mercury 6.2 are clearly not good
enough to perform this recovery at a 400-day timstep; better correctors
\citep{Wisdom06Correctors} or warmup \citep{SahaTremaine92} may be able to achieve this.

The convergence of our results at timesteps of 8 days or less, as well as the
agreement with two different non-symplectic integrators (including the
one used by Newman \etal), indicate that the IC perturbations described
by \citeN{SahaTremaine92} and \citeN{WisdomHolmanTouma96} are negligible in our smaller timestep cases.
Thus, using ``warmup'' \citep{SahaTremaine92} or higher-order symplectic
correctors will not substantially alter
our conclusions, although they might allow the same conclusions to be drawn
using larger timesteps.

With the exception of the results plotted in Figure \ref{fig:5Gy},
all of our simulations had a duration of 200 million years (My).  All of
them display an initial period of polynomial divergence, before the
appearance of exponential divergence (if any).  However, the duration
of initial polynomial divergence differs greatly across systems,
and has been observed by others \citep{LecarFranklinHolmanMurray01} to
last significantly longer that 200My; \citeN{GrazierNewmanHymanSharp05}
observed it to last the entire duration of their 800My simulation.  It
would be interesting to create a table like Table \ref{tab:chaosCounts},
but including a ``simulation duration'' dimension as well.  Certainly the
evidence hints that more systems make the ``switch'' from polynomial
to exponential divergence as the duration of the simulation increases.


Our physical model is very simplistic, accounting only for Newtonian
gravity between the Sun and Jovian planets.  Although we ignore many physical
effects which are known to effect the detailed motion of the planets
\citep{Laskar99,VaradiRunnegarGhil03}, it is unclear if such effects would
substantially alter the chaotic nature of solutions.  We at first believed
that the largest such effect ignored was solar mass loss.  Our first simulations
did not account for solar mass loss, which amounts to about one part in $10^7$
per million years \citep{Laskar99,Noerdlinger05}.  Since we find that perturbations
in position of that order can shift the system in-and-out of chaos, a naive
analysis might lead one to suspect that solar mass loss might shift the
planetary orbits in-and-out of resonance on a timesclase that is fast
compared to the Lyapunov time, thus smoothing out the sibling divergence.
We thus modified our model to include solar mass loss, but surprisingly it
made absolutely no observable difference to any of the figures presented
in this paper.  To ensure that we did not make an error, we simulated systems
with ever increasing mass loss until the Sun was losing 10\%
of its mass per 100My.  We noted that the planetary orbital semi-major axes
expanded significantly, as would be expected, but that the sibling divergences
did not change until mass loss was at a rate of about 1\% per 100My (1000 times
greater than in reality).  Thus, we conclude that solar mass loss also makes no
difference to our results.

\section*{Acknowledgements}
The author thanks Scott Tremaine, Norm Murray, Matt Holman, Philip Sharp,
and Bill Newman for helpful discussions and comments on the manuscript;
Norm Murray and Philip Sharp for sending me (and explaining) their ICs;
and Ferenc Varadi for giving me the source code to his most recent version
of {\tt NBI}.

\bibliography{ms}

\end{document}